\documentclass[twocolumn,amsmath,amssymb,prl,10pt,nofootinbib,superscriptaddress]{revtex4}
\usepackage{ graphicx, float,amsmath, amsmath, amssymb}
\usepackage[export]{adjustbox}
\usepackage[breaklinks, colorlinks, citecolor=blue]{hyperref}
\usepackage[labelsep=period]{caption}
\usepackage[normalem]{ulem}
\usepackage{mathtools}
\usepackage{siunitx}
\usepackage{soul}
\usepackage{physics}
\usepackage{minitoc}

\usepackage{bm}
\usepackage{xcolor}

\newcommand{\mubar}{\bar{\mu}}
\newcommand{\sums}{\sum\limits}

\definecolor{alizarin}{rgb}{0.82, 0.1, 0.26}

\def\be{\begin{equation}}
\def\ee{\end{equation}}
\def\bea{\begin{eqnarray}}
\def\eea{\end{eqnarray}}
\def\bse{\begin{subequations}}
\def\ese{\end{subequations}}

\begin{document}
\title{Cosmological implications of generalized holonomy corrections}

\author{Cyril Renevey}%
\affiliation{%
Laboratoire de Physique Subatomique et de Cosmologie, Universit\'e Grenoble-Alpes, CNRS/IN2P3\\
53, avenue des Martyrs, 38026 Grenoble cedex, France
}

\author{Killian Martineau}%
\affiliation{%
Laboratoire de Physique Subatomique et de Cosmologie, Universit\'e Grenoble-Alpes, CNRS/IN2P3\\
53, avenue des Martyrs, 38026 Grenoble cedex, France
}

\author{Aurélien Barrau}%
\affiliation{%
Laboratoire de Physique Subatomique et de Cosmologie, Universit\'e Grenoble-Alpes, CNRS/IN2P3\\
53, avenue des Martyrs, 38026 Grenoble cedex, France
}




\date{\today}
\begin{abstract} 
Most of the phenomenology of loop quantum gravity in the cosmological sector is based on the so-called holonomy correction to the Hamiltonian constraint. It straightforwardly modifies the Friedmann equations. In this work, we investigate the influence of corrections generalizing the one usually used in loop quantum cosmology. We find that a long enough inflation phase can be generated by purely quantum geometrical effects but we also underline the limitations of this scenario. In addition, we study the effects of generalized holonomy corrections on an inflationary phase generated by a massive scalar field. At the level of perturbations, we investigate in detail the consequences on the primordial scalar power spectrum. The results are actually quite general and can be used beyond the ``loop" framework.
\end{abstract}
\maketitle

\section{Introduction}

Loop quantum gravity (LQG) is a nonperturbative framework \cite{Ashtekar:2021kfp} providing a tentative quantization of general relativity (GR). It has been expressed both in the canonical form \cite{Rovelli:2004tv} and in a covariant way \cite{Rovelli:2014ssa}. As for all speculative theories the challenge is twofold. On the one hand, one has to check internal consistency. This is far from being a trivial requirement, especially in quantum gravity. From gauge issues to infrared corrections, quite a lot of questions remain -- at least partially -- open (see, {\it e.g.}, \cite{Kiefer:2007gns}). On the other hand, it is mandatory to face the outstanding question of phenomenological consequences \cite{Barrau:2013ula}. Making links with observations is the key missing ingredient for all attempts in quantum gravity, including string theory \cite{Quevedo:2016tbh}.\\

In this article, we address the question of the robustness of some predictions of LQG in the cosmological sector. Many different aspects have already been investigated, taking into account in particular (see, {\it e.g.} \cite{Ashtekar:2007em,lqc9,Agullo2,Diener:2014mia,Ashtekar:2015dja,Bolliet:2015bka,Alesci:2015nja,Martineau:2017sti,Gielen:2017eco})
\begin{itemize}
    \item the way initial conditions are set,
    \item the validity of the minisuperspace approximation,
    \item the backreaction effects,
    \item the deformation of the algebra of constraints,
    \item the inclusion of shear and curvature,
    \item the quantization of  operators associated with negative powers of the volume operator,
    \item the inclusion of effects inferred from quantum reduced loop gravity or group field theory,
    \item numerical results beyond the semiclassical approximation, etc.
\end{itemize}
Here, we tackle a different and somehow underestimated question: the consequences of a generalized holonomy correction. The point is {\it not} to invent what would be a superexotic theory, with new free parameters, to boost the phenomenological richness. Just the other way round, the aim is to investigate how reliable are the predictions made so far, taking into account implicit assumptions that went mostly unnoticed and may play an important role.\\

The issue of quantization ambiguities in this framework was pointed out in \cite{Perez:2005fn}. Those associated with the quantization of the connection-based holonomy variable might deeply influence the dynamics and constitute the subject of this article. New theoretical arguments are being given in \cite{Amadei:2022zwp}, while the present work focuses on potential observable effects. The question is especially important and meaningful when considered from a renormalization point of view.

In the following, the basics of loop quantum cosmology (LQC) are first briefly reminded. We then go into the details of generalized holonomy corrections. In the next section, we show that a long period of inflation can be generated using only a modified holonomy correction, without any massive scalar field. We also highlight the limits of such a model. The consequences of generalized holonomy corrections on both the background inflationary dynamics generated by a  massive inflaton field and the scalar primordial power spectrum are finally exposed.

\section{FLRW loop quantum cosmology}

In order to set the notations and remind the basics to the unfamiliar reader, we summarize the main ideas behind LQC. This also allows the article to be self-contained. In the fully constrained Ashtekar-Barbero formulation of GR, the canonical variables are
\begin{align}
    A^i_a \equiv \Gamma^i_a+\gamma K_a^i\quad\textrm{and}\quad E_i^a \equiv \frac{1}{2}\varepsilon^{abc}\varepsilon_{ijk}e^j_b e^k_c~,
\end{align}
where $\Gamma_a^i$ is the $su(2)$ spin connection, $\gamma$ the Barbero-Immirzi parameter, $K_a^i$ the extrinsic curvature, $\varepsilon^{abc}$ the totally anti-symmetric tensor, and $e^i_a$ the triads. In this work, we use $a,b,c$ as spacetime indices and $i,j,k$ as internal $su(2)$ algebra indices. Both sets run from 1 to 3. These canonical variables satisfy the relations
\begin{align}
    \left\{A^i_a(x),E^b_j(y)\right\}=\kappa \gamma\delta^i_j\delta^b_a\delta^3(x-y)~,
\end{align}
with $\kappa=8\pi G$. 

As GR is a fully constrained theory, its Hamiltonian is written as a sum of constraints, respectively called scalar, vector and Gauss constraints. When homogeneity is assumed, the scalar contribution $C_g$ is the only one to remain. The Hamiltonian can then be written using the lapse function $N$ as
\begin{align}
    C_g^N&=\int_\Sigma\dd x^3 N C_g
    \\
    &=\frac{1}{2\kappa}\int_\Sigma \dd x^3\frac{N}{\sqrt{q}}E^a_iE^b_j\left(\varepsilon^{ij}_kF^k_{ab}-2(1+\gamma^2)K^i_{[a}K^j_{b]}\right)~,
\end{align}
with $\Sigma$ a compact hypersurface.
In a homogeneous, isotropic, and flat space, the metric reduces to the form
\begin{align}
    \dd s^2=-\dd t^2+a^2(t)\delta_{ab}\dd x^a\dd x^b,
\end{align}
where the cosmic time is related to the $0$-coordinate by $\dd t=N\dd x^0$. To avoid divergent integrals and an ill-defined symplectic geometry, we perform the integration on an arbitrary cubic fiducial cell of comoving volume $V_0$. As the homogeneity assumption also implies that spatial derivatives vanish, the spin connection disappears and the canonical variables become simply
\begin{align}
    A^i_a(t) = \gamma \dot{a}(t) \delta^i_a \equiv \frac{c(t)}{V_0^{1/3}} \delta^i_a ~,
\end{align}

where the dot represents a derivative with respect to the cosmic time $t$, and 

\begin{align}
E^a_i(t) = a^2(t) \delta^a_i \equiv \frac{p(t)}{V_0^{2/3}} \delta^a_i~,
\end{align}


with the relation 
\begin{align}
    \{c,p\}=\frac{\kappa \gamma}{3}~.
\end{align}
Finally, the scalar constraint in this setting reduces to the simple form
\begin{align}
    C_g^N=-\frac{3}{\kappa \gamma^2}N\sqrt{p}c^2~.
\end{align}
The lapse function $N$ represents a gauge freedom.\\

In addition to the gravitational sector, we introduce a scalar field $\phi$ with an arbitrary potential $V(\phi)$ to investigate an early inflationary period. Using the canonical variables for the scalar field, namely $\phi$ and $\pi_\phi=p^{3/2}\dot{\phi}$, such that $\{\phi,\pi_\phi\}=1$, the total Hamiltonian describing the coupled system can be written as
\begin{align}
    C^N=C^N_g+C^N_m=N\left(-\frac{3}{\kappa \gamma^2}\sqrt{p}c^2+p^{3/2}\rho\right)~,
\end{align}
where $\rho=\pi_\phi^2/(2p^3)+V(\phi)$. The first Friedmann equation can easily be recovered using the evolution equation -- that is $\dot{p}=\{p,C^N\}$ -- with the choice $N=1$ and the Hamiltonian constraint. In its usual form, it is written as
\begin{align}
    H^2 \equiv \left(\frac{\dot{a}}{a}\right)^2=\frac{\kappa}{3}\rho~.
\end{align}

Up to now, we have simply recovered the usual GR result within a specific framework. An effective Hamiltonian including corrections from LQG is yet to be constructed. The first step toward canonical quantization with well-defined operators in the quantum theory is to rewrite the Hamiltonian constraint using the holonomy of the connection. In other words, instead of deriving the curvature operator $F_{ab}^i$ directly from the connection $c$, we use the holonomy $h_{\Box_{ij}}$ of the connection on a fiducial square curve $\Box_{ij}$ of length $\mu V_0^{1/3}$, with edges in the directions $i,j$. The holonomy measuring the extent to which the parallel transport of a vector around closed loops fails to preserve the transported vector, its form depends on the chosen $SU(2)$ representation for the parallel transport along the curves. This is known as the spin ambiguity. In standard LQC, the holonomy is calculated using the fundamental 2D representation of $SU(2)$, but the holonomy correction has been calculated for arbitrary representations in \cite{BenAchour:2016ajk}. In this section, we describe the procedure for the fundamental spin $1/2$ representation and we recall the procedure to follow for general corrections.\\

In the harmonic gauge, where $N=p^{3/2}$, the Hamiltonian constraint in terms of the curvature operator reads
\begin{align}
    C^h_g=-\frac{1}{2\kappa \gamma^2}p^2V_0^{2/3}\bar{e}^a_i \bar{e}^b_jF^k_{ab}~,
\end{align}
where the superscript $h$ stands for the harmonic gauge and $\bar{e}^a_i$ are the cotriads such that $q_{ab}=a^2(t)\bar{e}^i_a \bar{e}^j_b \delta_{ij}$. The holonomy corrected curvature operator in the fundamental representation is
\begin{align}
    F^k_{ab}=\lim\limits_{\mu\rightarrow 0}\frac{-2}{\mu^2 V_0^{2/3}}\Tr{h_{\Box_{ij}}\tau^k}\bar{e}^i_a \bar{e}^j_b~,
\end{align}
where $\tau^k$ are the generators of the $su(2)$ algebra and, in this case, are represented by the Pauli matrices. By taking the limit $\mu\rightarrow 0$, the usual definition of the curvature operator and of the Hamiltonian of standard GR are recovered. However, LQG teaches us that the lowest nonzero eigenvalue of the quantum area operator is $\lambda^2=4\sqrt{3}\pi \gamma l_{pl}^2$, hence taking the limit down to zero is in principle not allowed. One therefore chooses $\mu\rightarrow\bar{\mu}=\lambda/\sqrt{p}$ as an estimator of the smallest possible length for the edge of the square curve. The holonomy can then be calculated along a closed curve defined by
\begin{align}
    h_{\Box_{ij}}&=h_i\circ h_j\circ h_{-i}\circ h_{-j}~,
    \\
    \textrm{where}\quad h_{\epsilon i}:&=\exp{\epsilon\mu c \tau _i}~,
\end{align}
with $\epsilon=\pm 1$. Putting everything together, one obtains the Hamiltonian constraint of LQC coupled to a scalar field, that is
\begin{align}
    C^h=-\frac{3}{\kappa \gamma^2\bar{\mu}^2}p^2\sin^2(\mubar c)+p^3\rho~.
\end{align}

This modified Hamiltonian can be recovered from the Hamiltonian of GR using the substitution
\begin{align}
    c^2\rightarrow \frac{\sin^2(\mubar c)}{\mubar^2}~,\label{eq:lqc_holonomy_correction}
\end{align}
usually called ``the holonomy correction". Finally, it is possible to derive the modified Friedmann equation of LQC using $\dot{p}=\{p,C^h\}$, together with the Hamiltonian constraint $C^h=0$. This leads to:
\begin{align}
    H^2=\frac{\kappa}{3}\rho\left(1-\frac{\rho}{\rho_c}\right)~,\label{eq:friedmann_lqc}
\end{align}
with $\rho_c=3/(\kappa \gamma^2\lambda^2)$. The remarkable feature of this new  equation is the resolution of the Big Bang singularity. When $\rho\rightarrow\rho_c$ the Hubble parameter vanishes, as obvious from Eq. \eqref{eq:friedmann_lqc}, and a bounce occurs instead of the GR singularity\footnote{We however want to emphasize that contrary to what is often believed, a past singularity is {\it not} unavoidable in GR, even without exotic matter contents \cite{Barrau:2020nek,Renevey:2020zdj}.}. When choosing the usual value for the Barbero-Immirzi parameter $\gamma=0.2375$, the critical density is of the order of $\rho_c\approx 0.41\rho_{\text{Pl}}$.\\

The general case for the curvature operator, calculated using an arbitrary $d$-dimensional irreducible representation of $SU(2)$, was studied in \cite{Vandersloot:2005kh,Chiou:2009yx}. A new closed formula for the Hamiltonian of flat FLRW models regularized with arbitrary spins was found in \cite{BenAchour:2016ajk} and happens to be polynomial in the basic variables, which corresponds to well-defined operators in the quantum theory (taking also into account the inverse-volume corrections). The key-point lies in the fact that the computation in a representation of spin $j$ of the trace of an $SU(2)$ group element does not require the explicit knowledge of all its matrix elements and can be reduced to an expression involving only the trace in the fundamental representation and the class angle. The curvature operator can then be written as
\begin{align}
    F_{ab}^k=\frac{-3}{d(d^2-1)}\frac{1}{\mubar^2V_0^{2/3}}\frac{\sin^2(\mubar c)}{\sin \theta}\pdv{}{\theta}\left(\frac{\sin(d\cdot \theta)}{\sin\theta}\right)\varepsilon^k_{ij}\bar{e}^i_a \bar{e}^j_b~,\label{eq:general_curvature}
\end{align}
with
\begin{align}
\label{class angle 1}
\theta=\arccos\left(\cos(\mubar c)+\frac{1}{2}\sin^2(\mubar c)\right)~.
\end{align} 
This was derived with a technique quite similar to the one described previously for the holonomy regularization. Another technique to find the curvature operator, called connection regularization, can also be effectively considered \cite{BenAchour:2016ajk}. In this approach, a new definition of the curvature operator, only valid in homogeneous space, is used and the result for $F_{ab}^k$ is slightly different. In the literature, higher order holonomy corrections were also investigated in details \cite{Mielczarek:2008zz,Hrycyna:2008yu,Chiou:2009yx}. They arise when higher order terms in powers of $\mubar$, usually neglected, are taken into account in the expression for the regularized curvature. There could exist a link between these higher order holonomy corrections and the contribution of higher spin representations. However, it was shown in \cite{BenAchour:2016ajk} that these effects have actually very different physical consequences. At any order in holonomy corrections, a physical Hilbert space can be rigorously constructed and a complete family of Dirac observable can be identified \cite{Chiou:2009yx}.

It is therefore mandatory to understand the cosmological implications of more general holonomy corrections. 

\section{Cosmology with arbitrary holonomy corrections}

Instead of focusing on specific cases within the LQC framework with either different spin representations or higher order terms, we remain as general as possible for the expression for the holonomy correction. This can be studied by the substitution 
\begin{align}
    c^2\rightarrow g^2(c,p)~,\label{eq:general_holonomy_correction}
\end{align}
where $g(c,p)$ is an arbitrary function such that, in the low energy limit, 
standard GR is recovered, that is $g(c,p)\rightarrow c$. It should be emphasized that this is not only a way of taking into account the lessons from specific situations in LQC, but that this also make sense from a fully generic quantization ambiguity/renormalization viewpoint. Furthermore, in the totally constrained Hamiltonian formalism of GR, all the constraints are first class. We do not relax this requirement so that the evolution operator keeps the subspace of physical states invariant. As shown in \cite{Han:2017wmt}, keeping only first class constraints and following the usual Dirac prescriptions adds an extra condition on the function $g(c,p)$:
\begin{align}
    g(c,p)=\frac{1}{\mubar}f(b)~,\label{eq:anomaly_free_condition}
\end{align}
where $f(b)$ is an arbitrary function of $b=\mubar c$, which must behave as $f(b)\approx b$ at low energies ({\it i.e.} when $b\ll 1$). As opposed to \cite{Han:2017wmt}, we used the parameter $\lambda$ to respect the units of length of $g(c,p)$ and be consistent with LQC. Fortunately, this is fully compatible with the correction given in Eq. \eqref{eq:general_curvature}, ensuring that any holonomy modification coming from an arbitrary spin-representation will keep the algebra of constraints consistent. It is quite remarkable that the ``anomaly freedom" requirement (see \cite{Wu:2018mhg,eucl2,eucl3} for general considerations) allows one to sharpen the general expression, in a way precisely compatible with known corrections expected in the loop framework. The simple -- and mandatory -- fact that the evolution vector is asked to be parallel to the submanifold of constraints severely reduces the {\it a priori} freedom.\\

The modified equations of motion for the canonical variables $c$ and $p$ are calculated using Hamilton's equations. It is more natural to write them in terms of $p$ and $b$. Together with the Hamiltonian constraint, they take the form
\begin{align}
    \dot{b}&=-\frac{\lambda\kappa \gamma}{2}\rho(1+w)~,\label{eq:equation_for_c}
    \\
    \dot{p}&=\frac{2}{\gamma\lambda}p f(b)f'(b)~,\label{eq:equation_for_p}
    \\
    \frac{\rho}{\rho_c}&=f^2(b)~,\label{eq:hamiltonian_constraint}
\end{align}
where $f'(b)$ should be understood as $\dd f(b)/\dd b$, $w=P/\rho$ and $P=\pi_\phi^2/(2p^3)-V(\phi)$. To derive Eq. \eqref{eq:equation_for_c} we used the continuity equation
\begin{align}
    \dot{\rho}=-3\frac{\dot{p}}{2p}\rho(1+w)~.
\end{align}
A general modified Friedmann equation can be found using $H=\dot{p}/(2p)$ together with the constraint \eqref{eq:hamiltonian_constraint} and can be written as
\begin{align}
    H^2=\frac{\kappa}{3}\rho \left(f'(b)\right)^2~.\label{eq:general_friedmann}
\end{align}
Since, by construction, $f(b)\rightarrow b$ when $b\rightarrow 0$, we indeed recover the usual Friedmann equation $H^2=\kappa\rho/3$ in this limit. Let us now investigate the behaviour of the model starting in a regime where GR is valid and going backward in cosmic time $t$, toward a classical singularity associated with $\rho\rightarrow \infty$ in GR. Using the null energy condition, $w\geq -1$ together with Eqs.~\eqref{eq:equation_for_c} and \eqref{eq:equation_for_p} one can easily show that $\dd b/\dd t<0$ $\forall t$. Hence, if we start with $b>0$ in the GR regime, $b$ is always  positive and increasing when going backward in time. This is expected as, in the GR limit, the proportionality relation $\rho\propto b$ is satisfied and the density increases in the past direction. Furthermore, since one has $f(b)\approx b$ in the GR regime, the function $f$ is monotonic and strictly increasing with $b$ around $b=0$. In the case where there exists a local maximum $b_{bounce}>0$, implying $f'(b_{bounce})=0$, one can see with Eqs.~\eqref{eq:hamiltonian_constraint} and \eqref{eq:general_friedmann} that the density has to reach a critical value $\rho_{b}$, where the Hubble parameter vanishes. It is therefore meaningful to conclude that the Big Bang singularity is resolved by a bouncing scenario of geometrical origin if and only if the holonomy correction reaches a local maximum. If the function $f$ is strictly monotonic, two different scenarios have to be considered. Either $\lim\limits_{b\rightarrow \infty}f(b)=\infty$, in which case the singularity is not resolved, or $\lim\limits_{b\rightarrow \infty}f(b)<\infty$ and the situation is similar to eternal inflation where $\rho$ tends to a final constant value behaving as a positive cosmological constant.

\section{First remarks on inflation from the holonomy correction}
A natural question arising in this framework is to wonder whether it is possible to describe a long-lasting phase of inflation using only a modification of the holonomy correction without the need for a fluid satisfying the equation of state $w<-1/3$. In the usual LQC framework, the quantum geometrical super-inflation occurring after the bounce cannot account for more than a few e-folds and most of the known inflationary features are due to a hypothetical massive scalar field filling the Universe. Furthermore, one should also investigate if the inflation associated with generalized holonomies  could explain the quasi scale invariance of the power spectrum observed in the cosmological microwave background (CMB). The answer turns out to be positive. However, important drawbacks inherent to the construction will be mentioned in this section. We assume here that the content of the Universe is a massless scalar field, that is $w=1$ at all scales. We also restrict ourselves to holonomy corrections such that there exists a bounce so as to keep the huge benefit of the singularity resolution. \\

\begin{figure}
    \centering
    \includegraphics[width=0.42\textwidth]{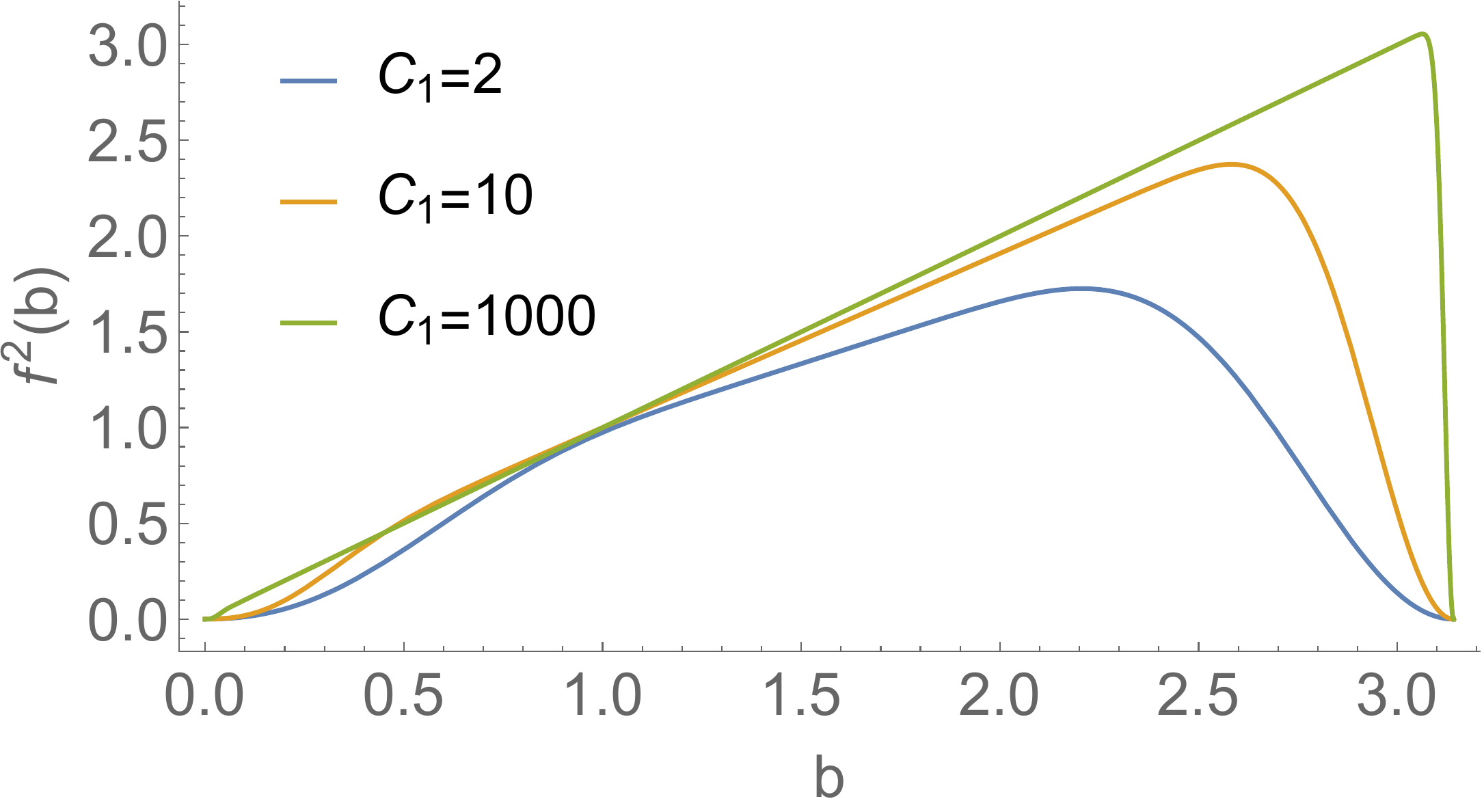}
    \caption{Holonomy corrections as described by Eq. \eqref{eq:correction_for_inflation} for different values of the parameter $C_1$.}
    \label{fig:inflation}
\end{figure}

First of all, the correction $f(b)$ has to be chosen so as to ensure an exponential growth of the scale factor $a$. This is achieved by $\dot{H}\simeq 0$: the Hubble parameter is nearly constant during inflation. From the modified Friedmann equation \eqref{eq:general_friedmann} and the Hamiltonian constraint $\eqref{eq:hamiltonian_constraint}$, one can easily conclude that the holonomy correction should take the following form:
\begin{align}
    f(b)\sim \sqrt{b}~.
\end{align}
We assume that $f(\pi)=0$ and call $b_{bi}$ the value of $b$ at the beginning of inflation. We set $b=b_{rh}$ at the reheating, when classical cosmology is recovered. It is therefore necessary to find a function $f(b)$ such that \textcolor{blue}{: i)} $f(b)\sim b$ around $b=0$, to recover GR, \textcolor{blue}{ii)} $f(b)\sim\sqrt{b}$ for $b_{rh}<b<b_{bi}$ during the inflationary period, and \textcolor{blue}{iii)} $f(b)$ reaches a maximum between $b_{bi}$ and $b=\pi$ to induce a bounce. In order to make explicit that such a function can be constructed, we give an arbitrary example:

\begin{align}
    f^2(b)&=(1+C_1 b)^{1-\alpha}\sin^2(b)\frac{1}{C_1+1}\sums_{n=0}^{C_1}\cos^{2 n}(b)
    \\
    &=\frac{(1+C_1 b)^{1-\alpha}}{1+C_1}\left(1-\cos^{2(C_1+1)}(b)\right)~,\label{eq:correction_for_inflation}
\end{align}

where $C_1$ can be chosen in accordance with the parameters $b_{bi}$ and $b_{rh}$ so as to obtain the desired inflationary duration. The parameter $\alpha$ can be chosen to make the power spectrum slightly red. To illustrate the behaviour of this holonomy correction, Eq.~\eqref{eq:correction_for_inflation} is plotted on Fig.~\ref{fig:inflation} for different values of $C_1$, assuming $\alpha=0$. The higher the value of $C_1$, the lower the value of $b$ at the end of inflation (denoted $b_{rh}$) and the closer $b_{bi}$ to $\pi$. Let us take a closer look at Eq. \eqref{eq:correction_for_inflation} in the case $C_1\gg 1$. If $b\gg 1/\sqrt{C_1}$, one indeed recovers $f^2(b)\sim b$, because the cosine term is small compared to $1$ and $C_1 b\gg 1$. On the other hand, when $b\ll 1/C_1$ one has $(1+C_1 b)\rightarrow 1$ while $(1-\cos^{2C_1}(b))\sim C_1 b^2$, leading to $f^2(b)\sim b^2$. However, in the intermediate case, $1/C_1\ll b\ll 1/\sqrt{C_1}$, $(1+C_1 b)\sim C_1 b$ while $(1-\cos^{2C_1}(b))$ still behaves as $C_1 b^2$, leading to $f^2(b)\sim b^3$.  Hence, the transition at reheating does not straightforwardly go from $f^2(b)\sim b$ to $f^2(b)\sim b^2$, but goes through transition phase, behaving as $f^2(b)\sim b^3$. This is not problematic for our purpose, but it is worth being pointed out. Equation~\eqref{eq:hamiltonian_constraint} shows that in order to have $N$ e-folds of inflation in a matter dominated universe, one needs $f^2(b_{rh})e^{3N}=f^2(b_{bi})\sim 1$. Therefore, if one chooses $C_1\sim e^{6N}$, the inflationary period ends when $b_{rh}\sim e^{-3N}\implies f^2(b_{rh}) \sim e^{-3N}$, leading to the desired $N$ e-folds. Then, the transition phase takes place while $e^{-6N}<b<e^{-3N}$ -- that can call reheating -- and finally we recover classical cosmology for $b<e^{-6N}$. Obviously, a quite strong fine-tuning is needed but solutions matching all the requirements can be found. We shall discuss this issue later.

Another potential problem to consider is related with the evolution of the energy density. In usual models of inflation, where the exponentially accelerating expansion is produced by a negative pressure fluid, the density of the latter stays roughly constant around (slightly above) the density of reheating $\rho\sim \rho_{rh}$. However, the density of radiation, dust or a massless scalar field evolve as $a^{-4}$, $a^{-3}$ and $a^{-6}$, respectively, thus increase exponentially during inflation as well (when thinking backward in time). If inflation is of quantum geometrical origin and not caused by the standard mechanism of a field slowly rolling on its potential, there is no reason to assume that the usual ``contents" are not present and possibly dominant. This means that if one requires a reheating around the GUT scale $T_{GUT}\sim 10^{15}$GeV and imposes roughly $N=65$ e-folds of exponential expansion, the energy at the beginning of inflation $T_{bi}$ would be vastly trans-Planckian. On the other hand, one can impose the maximum energy at the Planck scale $T_{bi}\sim T_{pl}$ and simply ask for more e-folds before the reheating than after, to solve the usual cosmological problems. However this translates into $T_{rh}\leq 1$ TeV. Although unusual, this value is not strictly ruled out by observations.\\

\begin{figure}
    \centering
    \includegraphics[width=0.4\textwidth]{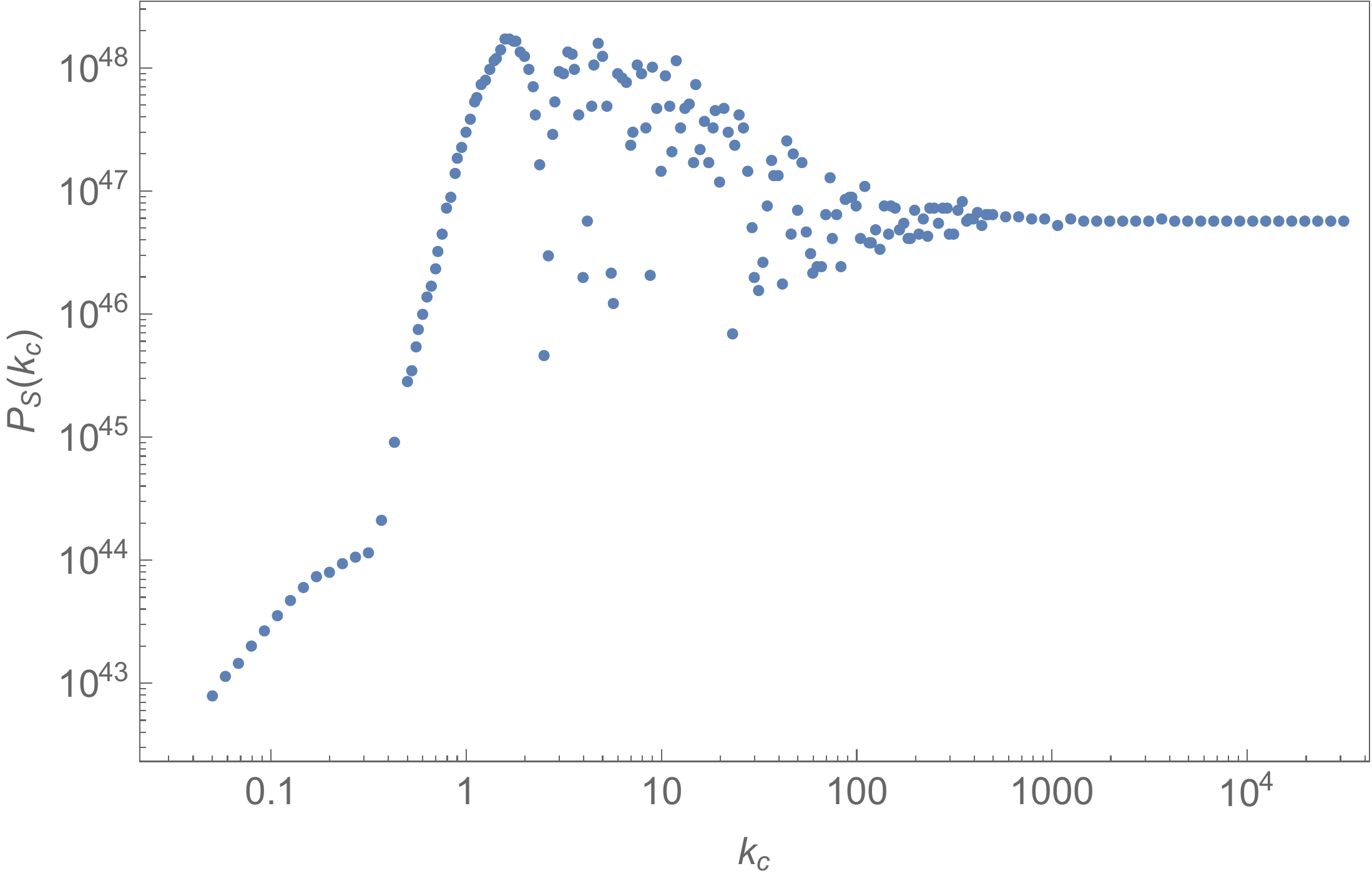}
    \caption{Scalar power spectrum obtained with a massless scalar field from a stage of ``geometrical" inflation entirely produced by the modified holonomy correction.}
    \label{fig:scalar_power_spectrum_massless}
\end{figure}

Finally, it is worth emphasizing that the scalar power spectrum, as observed in the CMB, cannot be easily recovered for other contents than a massive scalar field. To see this, let us consider a massless scalar field and a holonomy correction of the form \eqref{eq:correction_for_inflation} with a sufficiently long inflationary phase. In this example, we choose $T_{bi}\sim T_{pl}$ and roughly $N=35$ e-folds. The gauge invariant scalar perturbation can, in this context, be described by the Mukhanov-Sasaki (MS) variables $v_k(\eta)$ and $z(\eta)=\dot{\phi}a/H$, where $\eta$ is the conformal time, satisfying the equation
\begin{align}
    v_k''(\eta)+\left(k^2-\frac{z''(\eta)}{z(\eta)}\right) v_k (\eta) =0~.\label{eq:MS_equation}
\end{align}
It should be noticed that this classical equation is also the one fulfilled by perturbations in the dressed metric/hybrid quantization approach to LQC \cite{Agullo1,Fernandez-Mendez:2012poe}. 
The scalar power spectrum is obtained by
\begin{align}
    \mathcal{P}_S(k)=\left.\frac{k^3}{2\pi^2}\abs{\frac{v_k}{z}}^2\right|_{k=aH}\equiv \left.\frac{k^3}{2\pi^2}\abs{\mathcal{R}_k}^2\right|_{k=aH},
\end{align}
where, in the standard inflation theory, the curvature perturbation $\mathcal{R}_k$ is constant for super-Hubble modes $k\leq aH$. The scalar power spectrum therefore keeps a low amplitude 
matching the observed value. In the case of the massless scalar field with an inflationary stage due to the holonomy correction, one can show using Eq. \eqref{eq:MS_equation} that $\mathcal{R}_k$ is not constant for super-Hubble modes anymore. During the inflationary period, the Hubble parameter $H$ is constant, $a\propto e^{Ht}$ and $\dot{\phi}=2\sqrt{\rho}$, meaning that the second MS variable behaves as
\begin{align}
    z(a)\sim a^{-1/2(1+3w)}.
\end{align}
In the considered example $w=1$ but we keep the equation of state arbitrary so that the conclusion remains general. Rewriting the Mukhanov-Sasaki equation \eqref{eq:MS_equation} with respect to $a$ and taking the limit of large values of $a$, as one might expect for super-Hubble modes, we get
\begin{align}
    \dv[2]{v_k}{a}+\frac{1}{a}\dv{v_k}{a}-\frac{2}{a^2}v_k=0,
\end{align}
with solutions $v(a)\sim a$ and $v(a)\sim 1/a^2$. In standard inflation with a massive scalar field, where $w\approx -1$, choosing the Bunch-Davies vacuum comes down to selecting the behaviour $v(a)\sim a$ in quasi-de Sitter space, hence one recovers $\mathcal{R}_k=v_k/z\sim a^0$. In the case of a massless field ($w=1$), such a choice of vacuum leads to the behaviour $\mathcal{R}_k\sim a^3$ and therefore the scalar perturbations with super-Hubble modes exponentially increase during inflation. To illustrate this, the scalar power spectrum computed from Eq. (\ref{eq:MS_equation}) for an inflation induced by a holonomy correction with massless scalar field is shown in Fig.~\ref{fig:scalar_power_spectrum_massless}. As one can see, the spectrum is scale invariant in the UV and keeps its usual shape but the amplitude is meaningless. In principle, there might exist another vacuum selecting $v(a)\sim 1/a^2$ and one would recover $\mathcal{R}_k\sim a^0$. However, in the case of other matter contents, such as radiation ($w=1/3$) or dust ($w=0$), the scalar perturbations $\mathcal{R}_k$ cannot be frozen for $k\geq aH$ in geometric inflation. This is an important point, often forgotten, which should be taken into account in phenomenological studies.

It should however be emphasized that perturbations were, more than a decade ago, reunderstood in the framework of the effective theory of inflation \cite{Cheung:2007st}. This ``new" approach somehow disentangles the question of perturbations themselves from the question of the process which generates them. From this point of view, this work is anyway interesting as the perturbations can be treated even without any use of a scalar field. All that is needed is a background and an effective ``clock".

\section{Focus on the duration of inflation}

\begin{figure}
    \centering
    \includegraphics[width=0.4\textwidth]{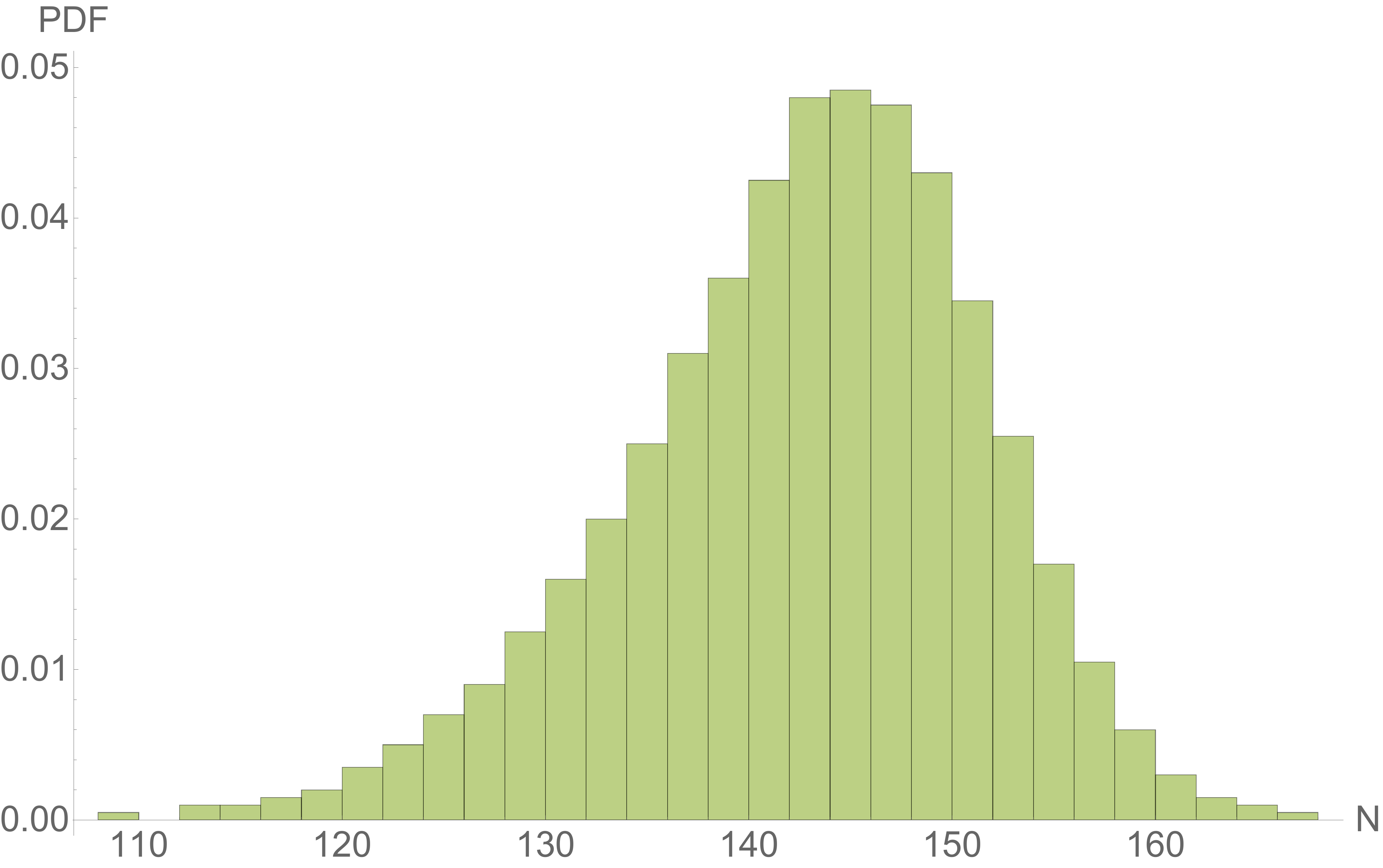}
    \caption{Probability distribution function of the number of e-folds of inflation for a quadratic potential with $m=1.2\cdot 10^{-6}$ in LQC.}
    \label{fig:PDFLQC}
\end{figure}

Detailed studies of the duration of inflation as predicted by LQC have already been made in \cite{Linsefors:2014tna,Martineau:2017sti,Bolliet:2017czc} for initial conditions set in the remote past. In the case of a massive scalar field playing the role of the inflaton, it was shown that the number of e-folds, as predicted by LQC, is around $N=140$. Interestingly, the number of e-folds varies only slightly with respect to most contingent parameters. This can easily be seen by studying trajectories in the potential energy $x(t)=\sqrt{V(\phi(t))/\rho_c}$ versus kinetic energy $y(t)=\sqrt{\dot{\phi}/(2\rho_c)}$ plane. The number $N=140$ corresponds to a quadratic potential $V(\phi)=m^2/2\phi^2$ with a mass parameter $m\sim 1.2\cdot  10^{-6}$. For other potentials, the number of e-folds can be different, but remains of this order of magnitude as long as one deals with confining potentials. Although anisotropies can slightly decrease this value, the duration of inflation is a robust property of the background in LQC (a different proposal was however suggested in \cite{Ashtekar:2009mm} but relies on conditions set at the bounce, a path that we do not follow here). The fact that this number is way smaller than na\"{\i}ve expectations is a key feature of bouncing scenarios. One can therefore naturally wonder if this prediction changes for different shapes of the holonomy correction. The set of initial prebounce conditions $(x_0,y_0)$ is described using a phase parameter $\delta\in[0,2\pi[$ such that
\begin{align}
    x_0=\sqrt{\frac{\rho_0}{\rho_c}}\cos{\delta}\quad \textrm{and}\quad y_0=\sqrt{\frac{\rho_0}{\rho_c}}\sin{\delta},
\end{align}
where $\rho_0$ is the initial energy density and defines how far away in the past initial conditions are set. It has been checked that the chosen distribution for $\delta$ is conserved over time. For completeness, we reproduce the probability distribution function (PDF) using the settings of \cite{Linsefors:2014tna} in Fig.\ref{fig:PDFLQC} with a uniform distribution of the phase parameter $\delta$. In the following, the mean value of the number of e-folds for different holonomy corrections is calculated using such a PDF.

In this section, we investigate the effect of the form of the holonomy correction on the duration of inflation. For this purpose, we follow the strategy described in \cite{Martineau:2017sti,Bolliet:2017czc}:  we assume a uniformly distributed set of initial conditions for $\delta$ and the initial time $t=0$ sufficiently far in the contracting branch so as to remain away from the quantum gravitational regime. In order to stay conservative about the holonomy correction shape, we study different general properties such as an asymmetry, an increase of the maximum of $f^2$, and a flattening of the correction on the length of inflation. Overall, we want to keep the asymptotic behaviour of LQC, {\it i.e.} $f(0+\delta b)=f(\pi+\delta b)=\delta b+\mathcal{O}(\delta b^2)$, hence we choose a correction of the form 
\begin{align}
    f^2(b)=\sin^2(b)\left(1+A_1 b^{n_1}(b-\pi)^{n_2}\right),
\end{align}
where $n_i\geq 1$, $i=1,2$ and $A_1\geq 0$, or
\begin{align}
    f^2(b)=\sin^2(b)\frac{1}{C_1+1}\sums_{n=0}^{C_1}\cos^{2 n}(b),
\end{align}
where $C_1\geq 1$. Such parametrizations ensure the desired behaviour at the fix points $b=0,\pi$. 

\begin{figure}
    \centering
    \includegraphics[width=0.5\textwidth]{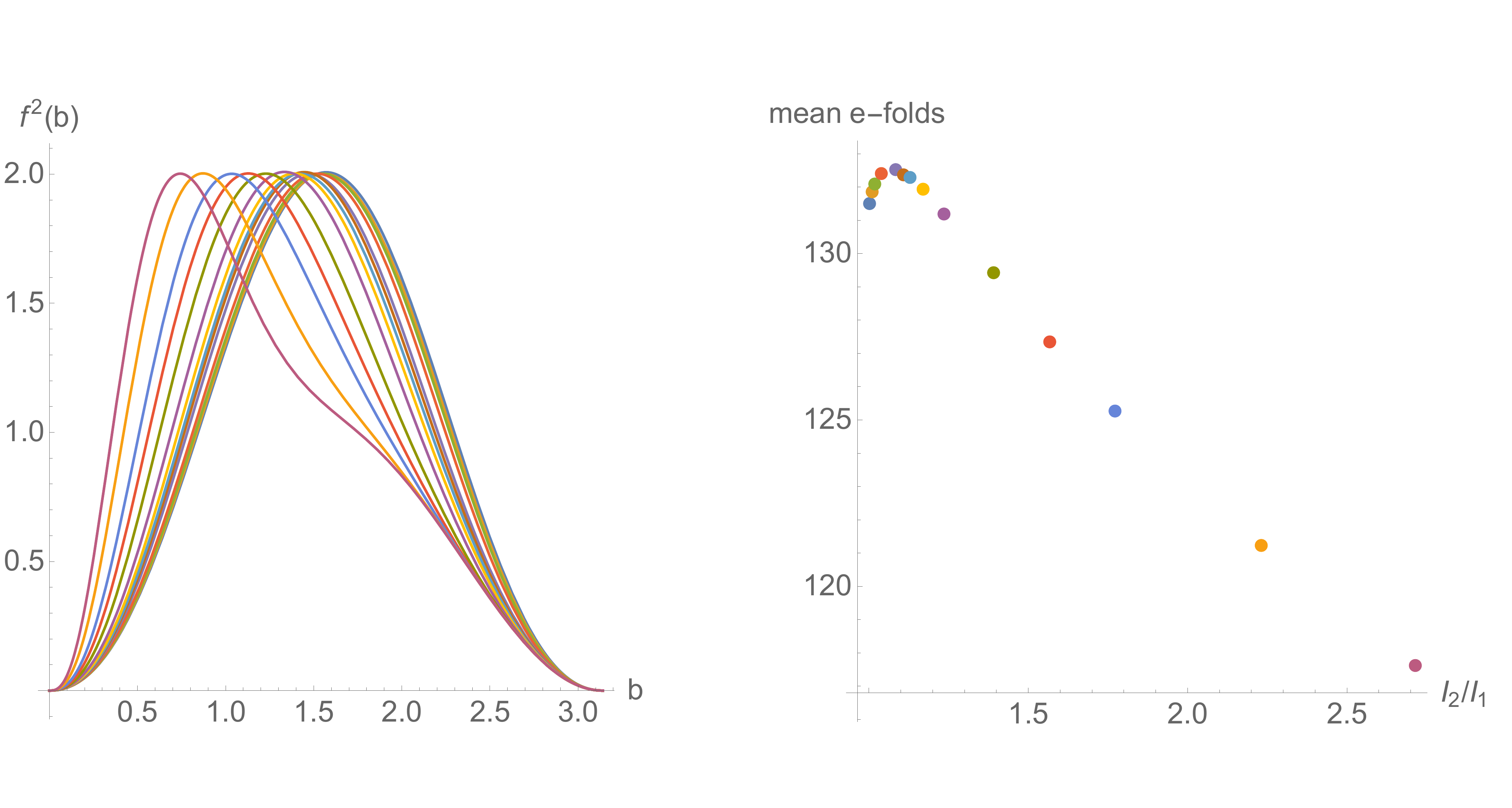}
    \caption{{\it Left}: asymmetric holonomy corrections. {\it Right}: mean value of the PDF of the number of inflationary e-folds.}
    \label{fig:asymmetries_left}
\end{figure}

\begin{figure}
    \centering
    \includegraphics[width=0.5\textwidth]{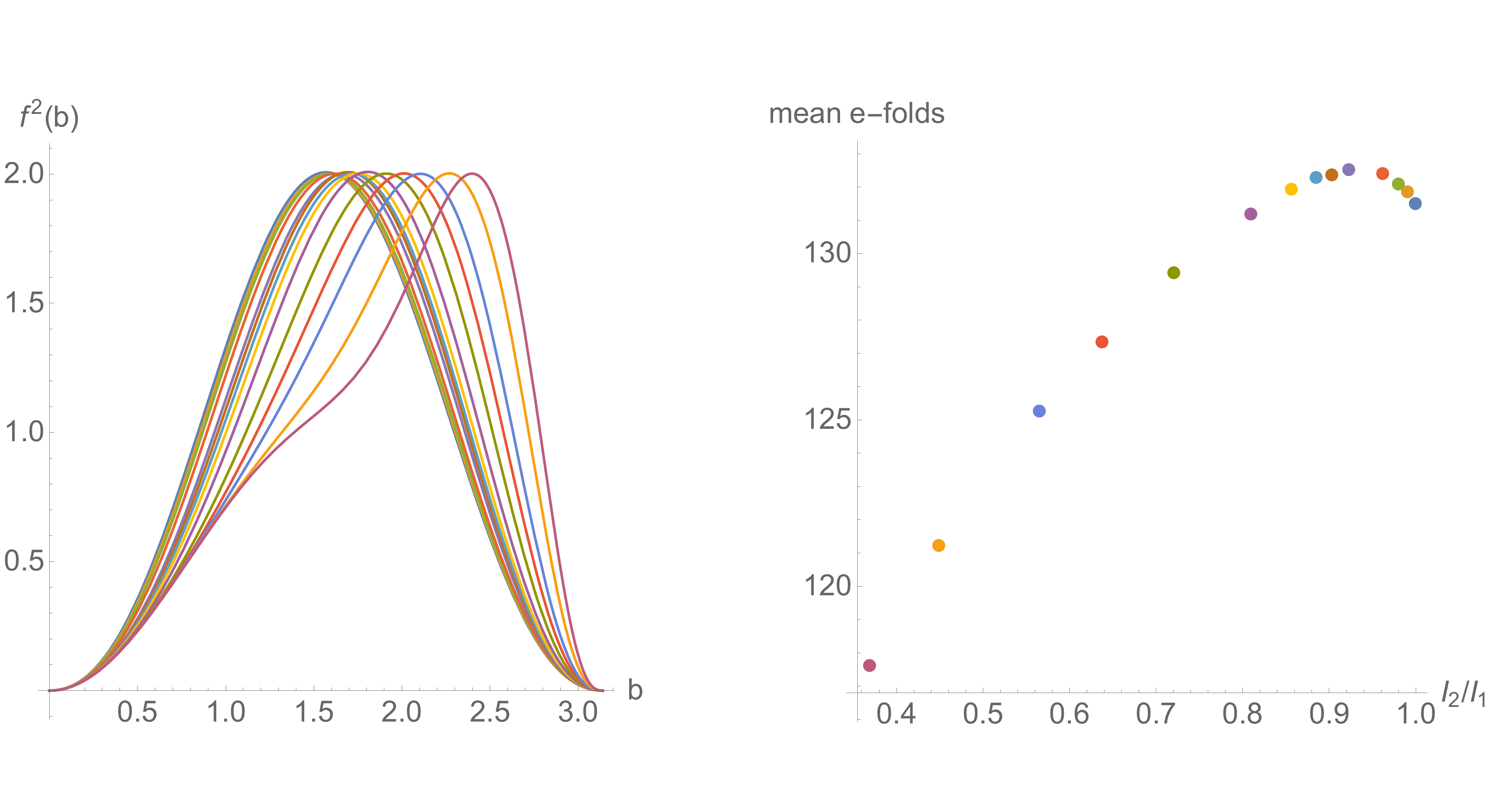}
    \caption{{\it Left}: asymmetric holonomy corrections. {\it Right}: mean value of the PDF of the number of inflationary e-folds.}
    \label{fig:asymmetries_right}
\end{figure}

The results of the numerical simulations for different left and right asymmetries are shown in Figs. \ref{fig:asymmetries_left} and \ref{fig:asymmetries_right}. The left panels represent the chosen holonomy corrections whereas the right panels display the mean value of the PDF of the number of e-folds when scanning the full range of initial phases. It should be pointed out that the PDF is narrow enough ($\sigma \sim 10$ e-folds) so that its first moment gives the relevant information. The strength of the asymmetry is measured by the ratio $I_2/I_1$ between the integral of $f^2(b)$ on $[0,b_{max}]$ and the integral on $[b_{max},\pi]$, where $b_{max}$ is the value that maximizes $f^2(b)$. The general trend is a decrease of the number of e-folds (although this is not true for very small deformations).

\begin{figure}
    \centering
    \includegraphics[width=0.5\textwidth]{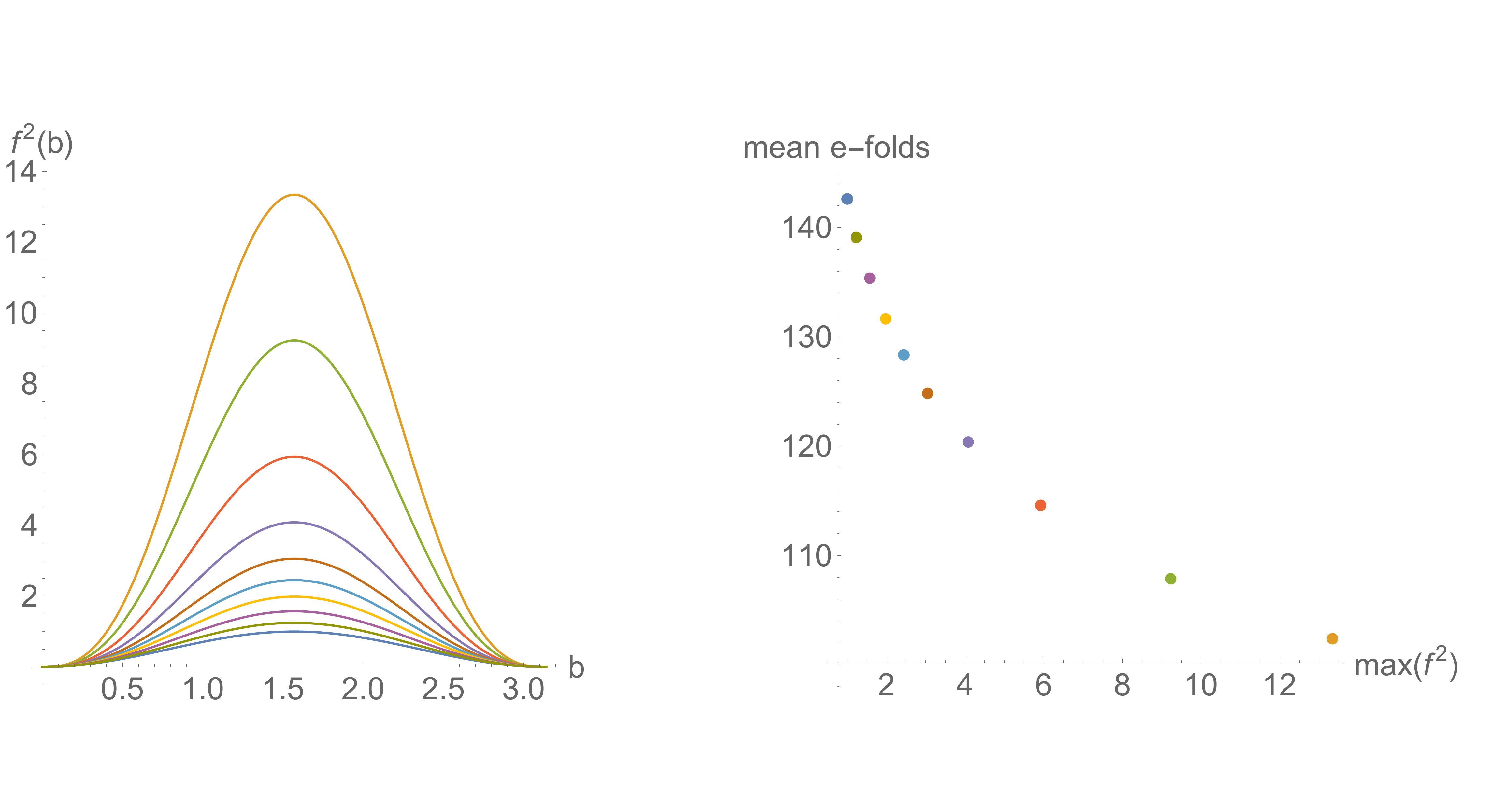}
    \caption{{\it Left}: amplitude-varying holonomy corrections. {\it Right}: mean value of the PDF of the number of inflationary e-folds.}
    \label{fig:maxf}
\end{figure}

\begin{figure}
    \centering
    \includegraphics[width=0.5\textwidth]{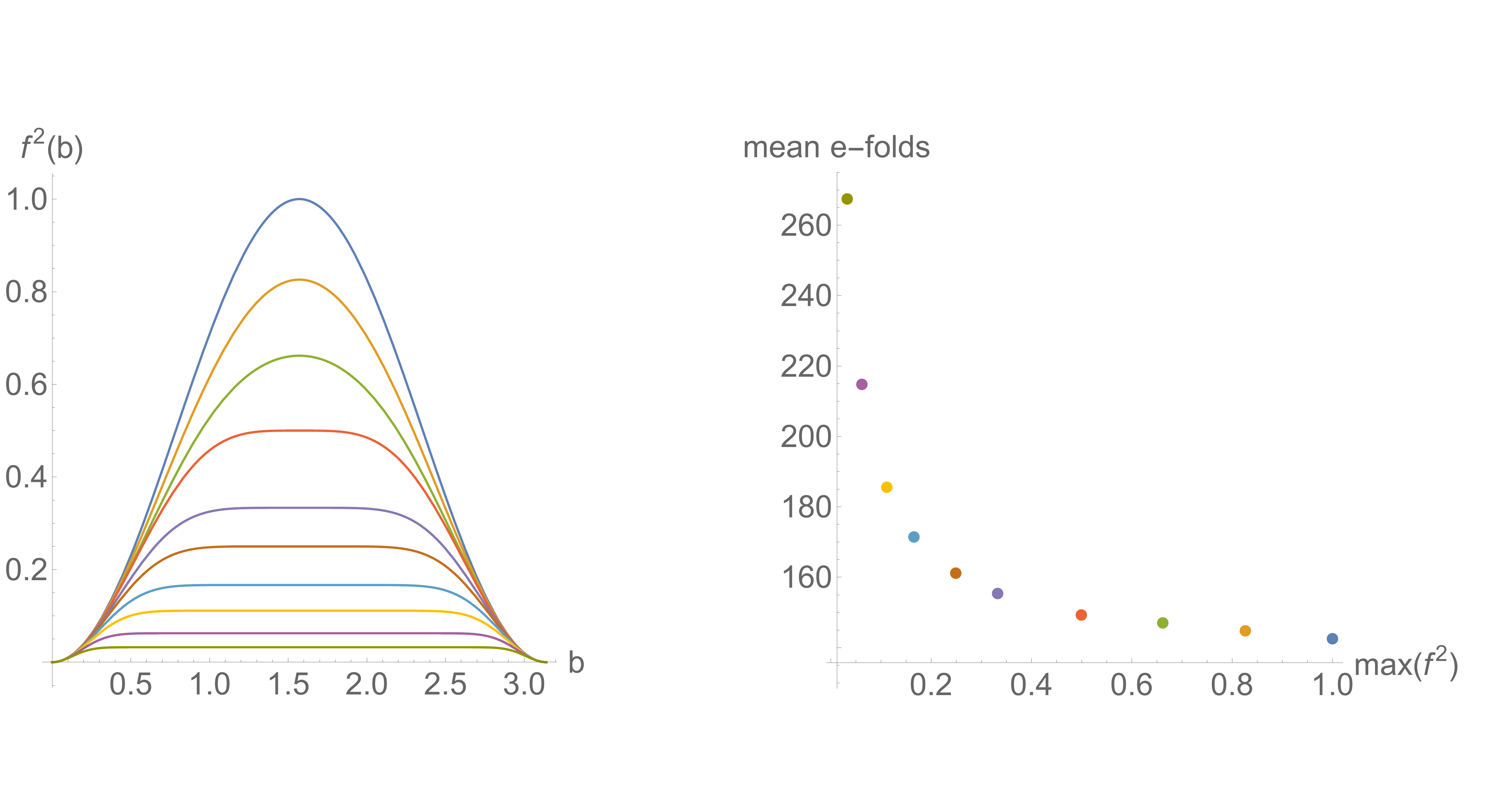}
    \caption{{\it Left}: plateau-like holonomy corrections. {\it Right}: mean value of the PDF of the number of inflationary e-folds.}
    \label{fig:plateau}
\end{figure}

\begin{figure}
    \centering
    \includegraphics[width=0.5\textwidth]{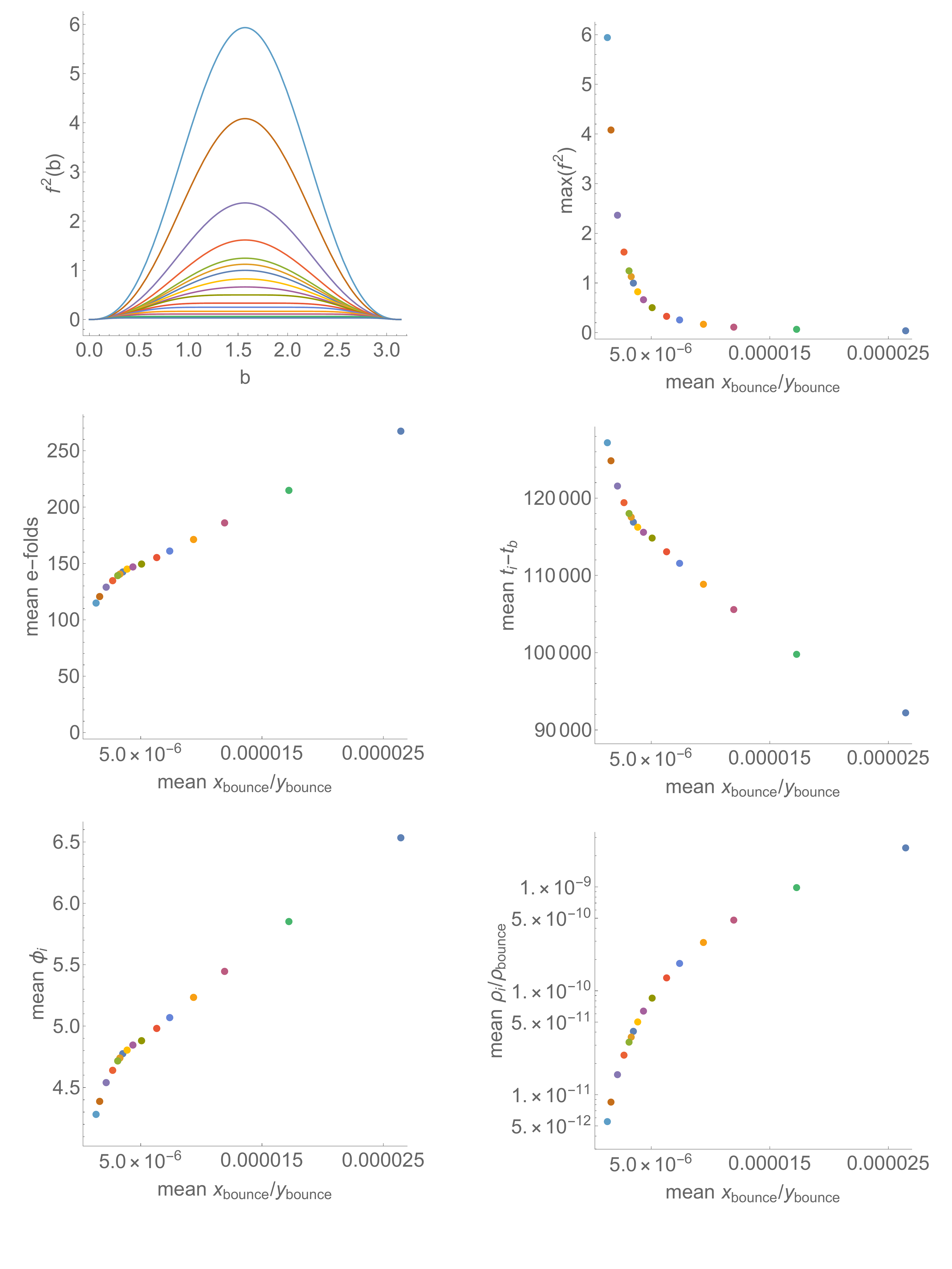}
    \caption{{\it Top left}: amplitude-varying holonomy corrections. {\it Top right}: maximum of $f^2(b)$ wrt. to the ratio of potential to kinetic energy of the scalar field at the bounce, $x(t_b)/y(t_b)$. {\it Middle left}: mean value of the PDF of the number of inflationary e-folds wrt. $x(t_b)/y(t_b)$. {\it Middle right}: mean value of the elapsed time between the bounce and inflation wrt. $x(t_b)/y(t_b)$. {\it Bottom left}: mean value of the scalar field at the beginning of inflation wrt. $x(t_b)/y(t_b)$. {\it Bottom right}: mean value of the ratio of the density at the beginning of inflation over the one at the bounce wrt. $x(t_b)/y(t_b)$.}
    \label{fig:full_ana}
\end{figure}

The results of the simulations for different amplitudes and plateaulike functions are displayed in Figs. \ref{fig:maxf} and \ref{fig:plateau}, respectively. In general, increasing the energy density at the bounce decreases the number of e-folds and flattening the correction (or reducing the energy of the bounce) increases the length of inflation. Since the bounce energy density grows up with $\max{f^2}$, as shown by Eq.(\ref{eq:hamiltonian_constraint}), this result seems counter-intuitive at first sight. It is true that the inflation duration generally increases with the bounce energy scale, but this is not the only important ingredient relevant for the number of inflationary e-folds. Another key parameter is the ratio between potential and kinetic energies at the bounce $x(t_b)/y(t_b)$. For kinetic energy dominated bounces, more time will indeed be required before the potential energy finally dominates and the slow-roll conditions are met. This leads to a lower inflation energy scale than for potential energy dominated bounces. Thus, bounces dominated by kinetic  energy will actually lead to shorter inflation phases. This is the reason why the PDF of the number of inflationary e-folds in LQC is peaked around low values of $N$, close to the experimental lower bound. Setting initial conditions in the remote past of the contracting branch with a flat PDF on the initial phase of the scalar field leads to trajectories with kinetic energy dominated bounces, with ratios $x(t_b)/y(t_b)$ typically of order $10^{-6}$. In the present study, the higher $\max{f^2}$ , the more kinetic energy dominated the bounce is. This  phenomenon is shown in Fig.\ref{fig:full_ana}. First, this indeed confirms in the top right and middle left plots, representing respectively $\max{f^2}$ wrt. $x(t_b)/y(t_b)$ and the number of e-folds wrt. $x(t_b)/y(t_b)$, that $\max{f^2}$ has a significant effect on this ratio, which in turn affects the number of e-folds as expected. Using the middle and bottom right plots, one can see that it takes longer for a  kinetic energy dominated bounce to start the inflationary period, leading to a lower energy density at inflation, thus lowering the number of e-folds. 

There are several lessons to be learnt from this analysis. There is an obvious loss of predictivity associated with a possible generalization of the holonomy correction. This can hardly come as a surprise. Whether this should be considered as good or bad news depends on the point of view. This obviously makes the theory less easily falsifiable. But this also opens up interesting possibilities. It shows that desired cosmological behaviours can be obtained by purely geometrical effects, relaxing the need for exotic matter contents. This also means that, in principle, a good knowledge of the cosmological dynamics can severely constrain the shape of the holonomy correction. In particular, the minimum number of e-folds $N>60$ required to solve the horizon problem sets a constraint on the maximum of $f^2(b)$ as $N$ decreases when the maximum density at the bounce increases. Most arbitrary shapes are problematic. It should however be underlined that as long as one deals with moderate corrections to the usual $\sin^2(\mubar c)/\mubar^2$, the effects on the background dynamics remain quite small and the core of the known LQC predictions remains valid.

The ambiguities considered in this work do {\it not}, by themselves, reveal a theoretical failure. So as to regularize quantum operators associated with nonlinear functionals of the fundamental fields, one relies on the diffeomorphism invariant prescription of ``point-splitting" \cite{Perez:2005fn}. It happens to be that the regulator can be removed without encountering UV divergences. One is then left with a well defined quantum Hamiltonian constraint, at the price of having many different quantum theories. This is obviously reminiscent of the usual problem of renormalization of quantum fields: the correct theory must be fixed by the  renormalization conditions.

\section{Effects on the primordial power spectra}


We now analyze the effects of a modified holonomy correction on the scalar power spectrum when considering an inflationary period generated by a massive scalar field of mass $m=1.2\times 10^{-6} ~m_{\text{Pl}}$. We keep the form of the Mukhanov-Sasaki equation for the scalar perturbations unchanged. This allows to specifically investigate the effects of the modification of the background dynamics, due to the new holonomy correction, on the shape of the spectra, independently of possible modifications of the perturbations propagation equation (which is anyway beyond the scope of this study). 

\begin{figure}
    \centering
    \includegraphics[width=0.5\textwidth]{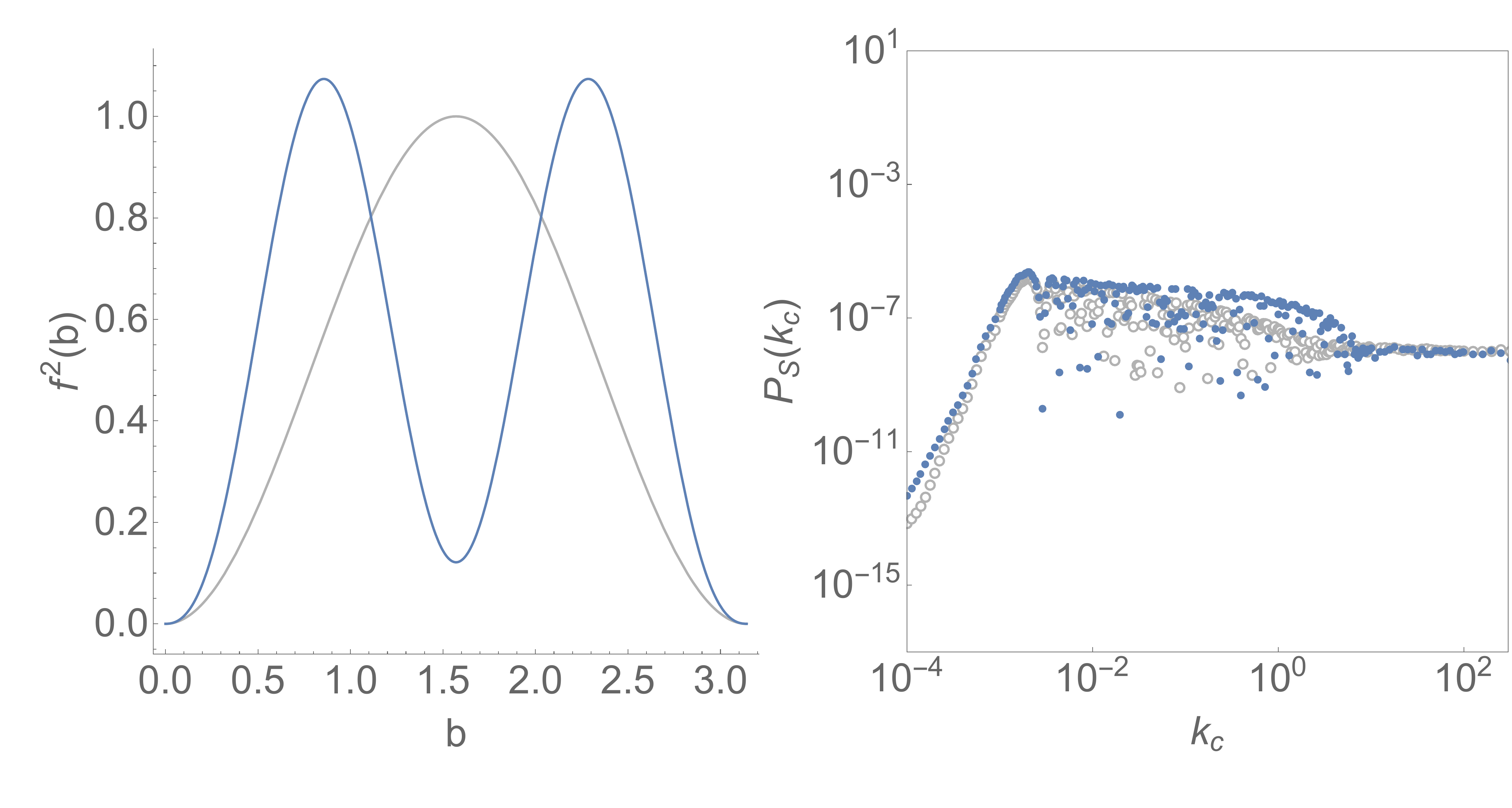}
    \caption{{\it Left}: holonomy corrections for standard LQC (in gray) and for a double bounce (in blue). {\it Right}: comparison between the associated scalar power spectra (standard LQC in gray and with modified holonomy correction in blue).}
    \label{fig:spectre_double}
\end{figure}
\begin{figure}
    \centering
    \includegraphics[width=0.5\textwidth]{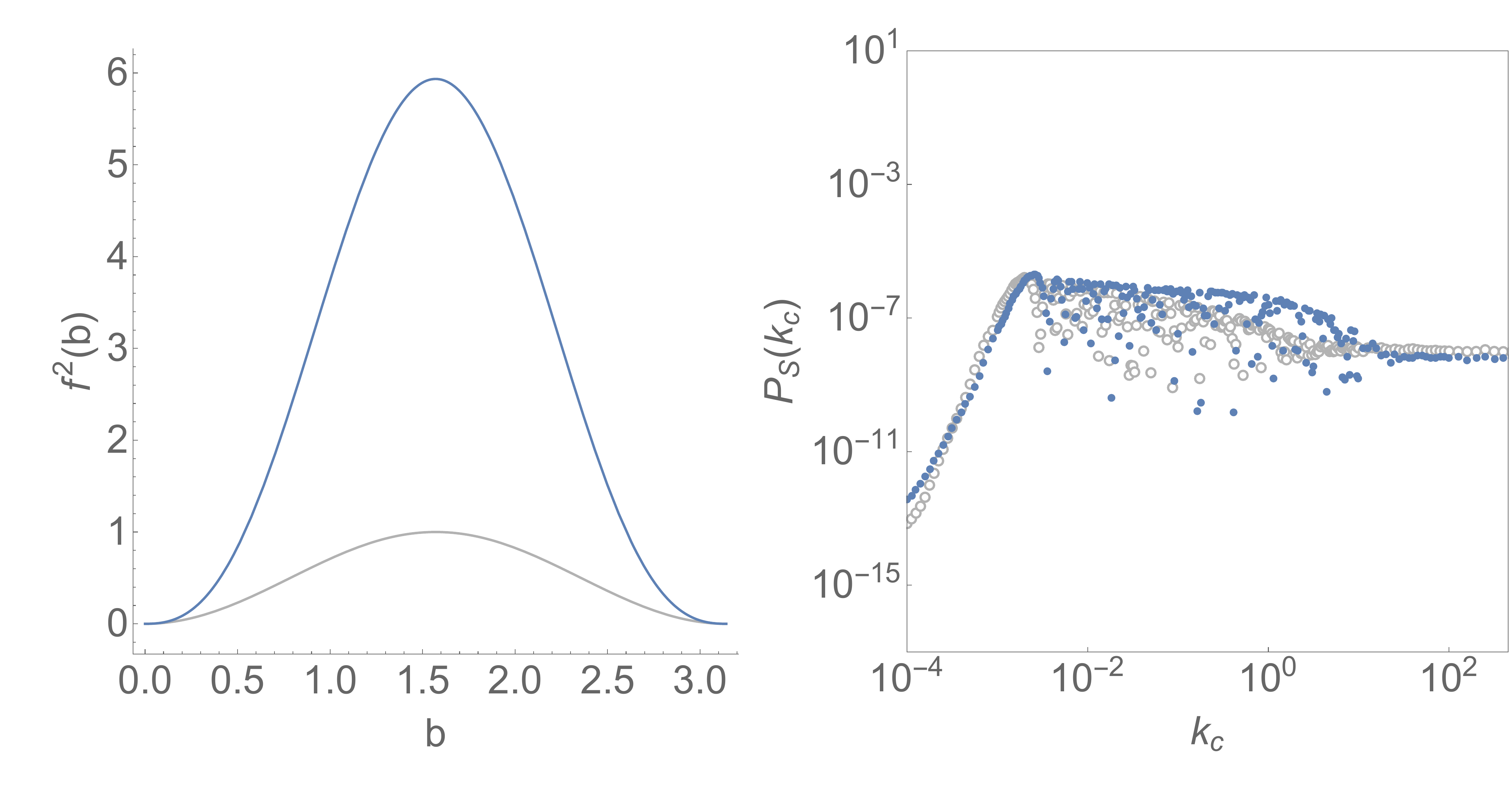}
    \caption{{\it Left}: holonomy corrections for standard LQC (in gray) and for a higher energy bounce (in blue). {\it Right}: comparison between the associated scalar power spectra (standard LQC in gray and with modified holonomy correction in blue).}
    \label{fig:spectre_maxf}
\end{figure}

Figures \ref{fig:spectre_double} and \ref{fig:spectre_maxf} display comparisons of the usual LQC spectrum (in grey) with spectra obtained using new background dynamics due to modified holonomy corrections (in blue). Figures \ref{fig:spectre_double} represents a double bounce while a single bounce at higher energy is shown in Fig. \ref{fig:spectre_maxf}. The new  features in the spectra corresponding to the modified dynamics are hardly distinguishable from the usual spectrum. The spectra remain (almost) scale invariant and consistent with CMB data in the so-called ultraviolet region, that is the region corresponding to the highest presented wave numbers (and even higher values of $k_c$), in which the observable scales are located for almost all the parameter space. The modifications to $f^2(b)$ only impact the details of the oscillations in the intermediate regime. This was expected as those oscillations are mainly associated with the bounce and their details depend on the detailed dynamics at this time. The ultraviolet regime, \textit{i.e} the one of interest for comparison with data, is mostly independent of the bounce dynamics as the wavenumbers of the corresponding modes are much greater than the potential $z''/z$ during the bounce phase. In other words, those modes do not even feel the presence of the bounce(s) It is however worth emphasizing that, as highlighted in \cite{Martineau:2018isp}, the position of the oscillations is dictated by the value of the potential $z''/z$ evaluated at the bounce. Thus, sharper bounces would lead to oscillations in the spectra located further away in the ultraviolet regime. It is also worth to underline that modifications of $f^2(b)$ around $b \approx 0$ or $\pi$ would modify the low energy behavior and have a deeper impact on the spectra. 

\begin{figure}
    \centering
    \includegraphics[width=0.5\textwidth]{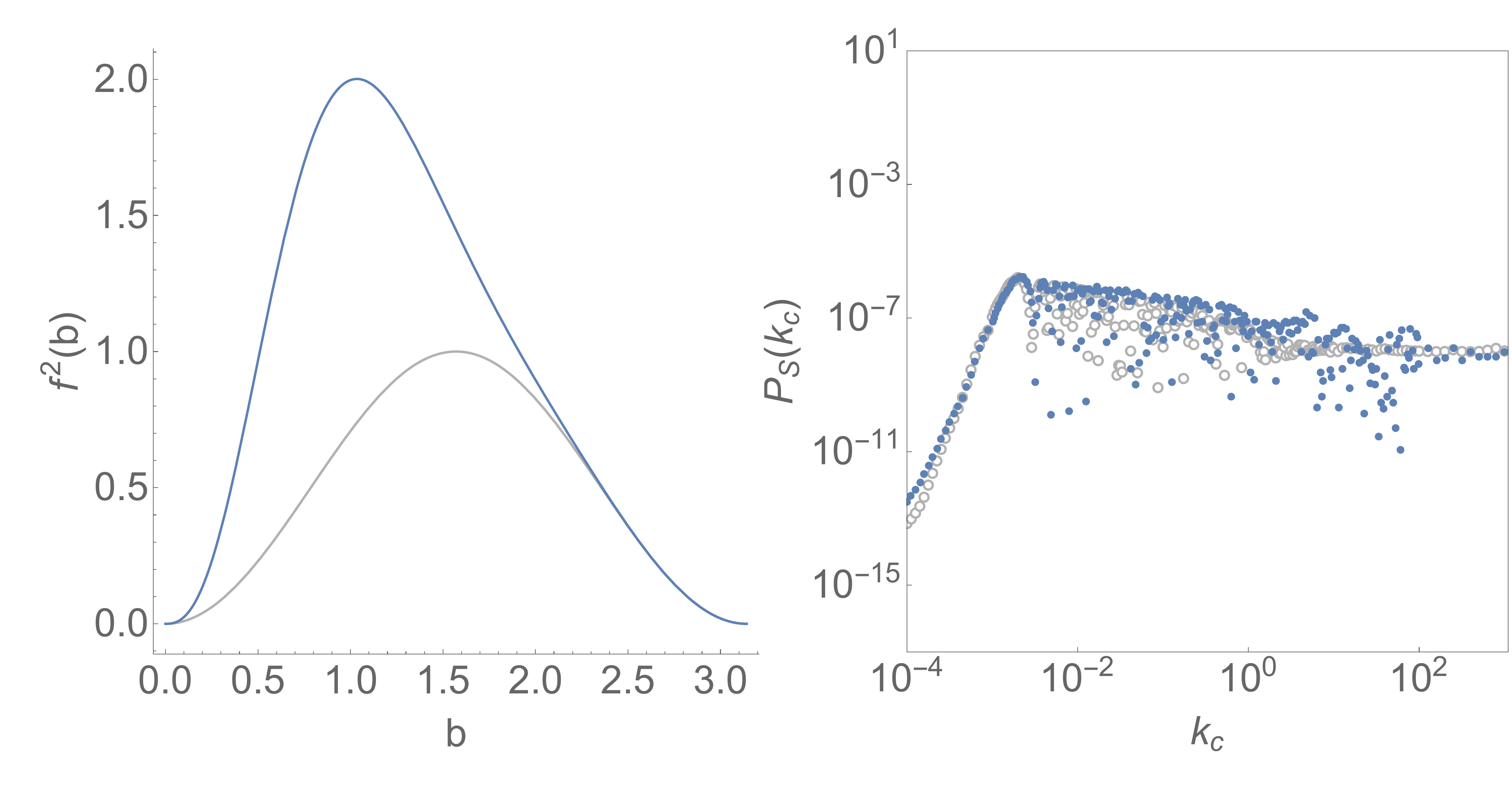}
    \caption{{\it Left}: holonomy corrections for standard LQC (in grey) and for a left asymetric bounce (in blue). {\it Right}: comparison between the associated scalar power spectra (standard LQC in grey and with modified holonomy correction in blue).}
    \label{fig:spectre_left}
\end{figure}

\begin{figure}
    \centering
    \includegraphics[width=0.5\textwidth]{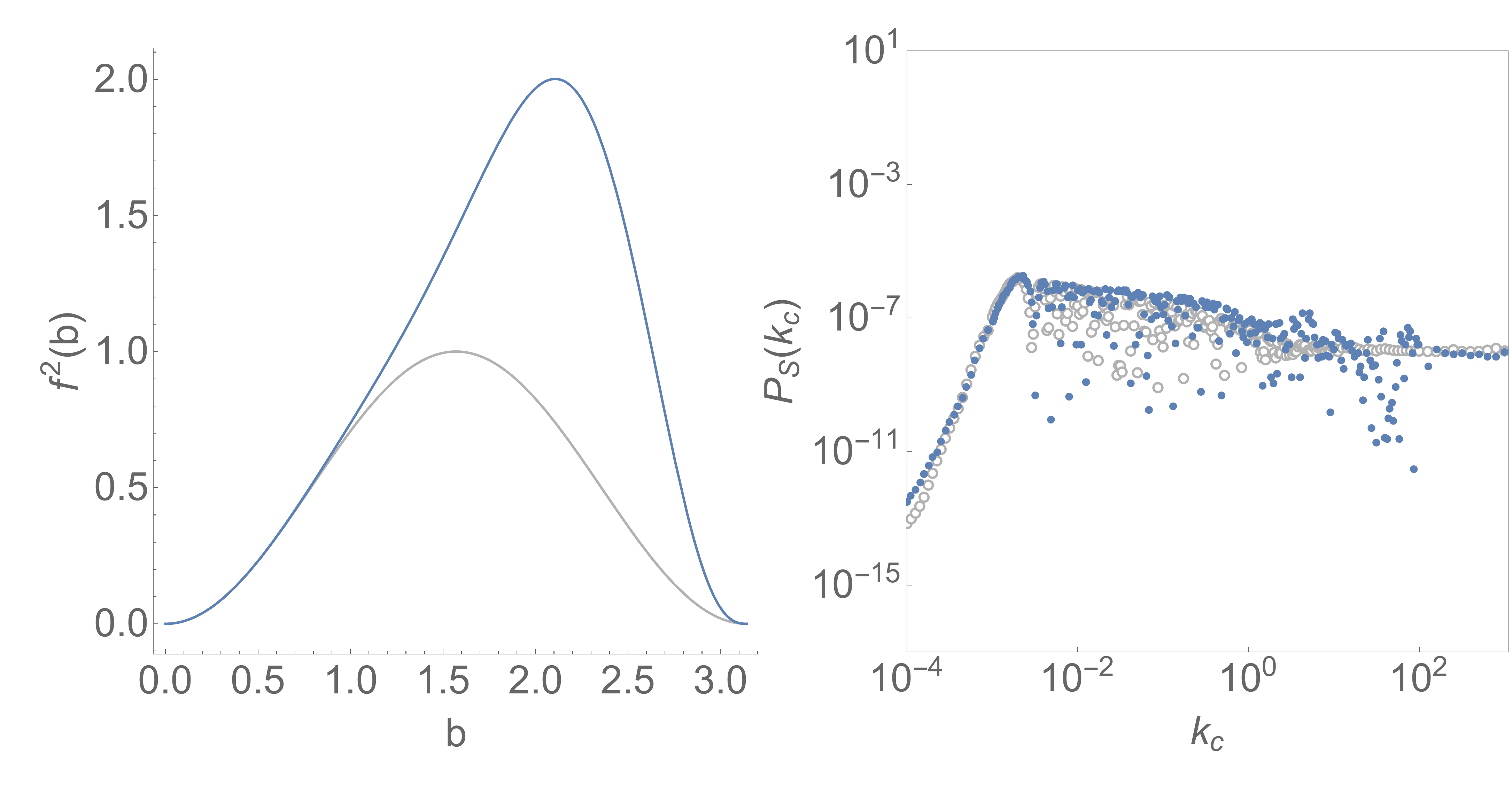}
    \caption{{\it Left}: holonomy corrections for standard LQC (in grey) and for a right asymetric bounce (in blue). {\it Right}: comparison between the associated scalar power spectra (standard LQC in grey and with modified holonomy correction in blue).}
    \label{fig:spectre_right}
\end{figure}

We have previously seen that an asymmetry (in $b$) of the holonomy correction can have a significant influence on the evolution of the inflaton in the high energy regime. It is therefore important to estimate the effects of asymmetries in $f^2(b)$ on the spectra. Associated results are shown in Figs. \ref{fig:spectre_left} and \ref{fig:spectre_right}.
The main noticeable effect is an extension of the oscillatory regime toward higher wave numbers.  This extension is however (for all the functions tested) not sufficient to be game changing for the noticeable features of primordial power spectra in loop inspired cosmologies. All usual conclusions still hold. This effect only makes a possible detection/invalidation of the bounce in CMB data slightly more probable as it decreases a little the level of fine tuning required to bring the window of comoving wave numbers associated with CMB observations toward the oscillatory regime.


It should be underlined that the situation might be totally different in the case of the Deformed Algebra (DA) approach to perturbations, where the equation of propagation of perturbations is usually given by\cite{Barrau:2018gyz,Schander:2015eja}:

\begin{equation}
v_k''(\eta)+\left(\Omega(\eta) k^2-\frac{z''(\eta)}{z(\eta)}\right) v_k (\eta) =0 ~,
\label{Eq. perturbations DA}
\end{equation}

with $\Omega = 1 - 2 \rho / \rho_c$. The change of sign of $\Omega$ at high energies can be interpreted as a switch from a Lorentzian to an Euclidean geometry \cite{Bojowald:2015gra}. The apparition of this $\Omega$ function in the MS equation comes from its presence in the anomaly-free algebra of constraints, more precisely in the Poisson bracket between scalar constraints\cite{eucl3}:

\begin{equation}
\left\lbrace S^{\text{a.f}}[M], S^{\text{a.f}}[N] \right\rbrace \approx \Omega D^\text{a.f} \left[ q^{ab} \left( M \partial_b N - N \partial_b M \right) \right],
\end{equation}

where $S$ and $D$ correspond respectively to the scalar and diffeomorphism constraints, the label a.f stands for "anomaly-free", and the $\approx$ symbol means that this relation is satisfied on the hypersurface of constraints. But the $\Omega$ function appears in this algebra as a consequence of the modification of the kinetic term in the scalar constraint (by the holonomy correction, that is Eq.(\ref{eq:lqc_holonomy_correction}) in the usual case). If we assume a 1+1 dimensional toy model then $\Omega = (1/2) d^2 f^2(b) / db^2$, which, in the case $f^2(b) = \sin^2(b)$, and using the Hamiltonian constraint, gives back the previous expression $\Omega = \cos(2 b) = 1 - 2 \rho / \rho_c$. It is thus clear that modifying the form of $f(b)$ would unavoidably modify the expression of $\Omega$ and therefore the Mukhanov-Sasaki equation. As a consequence, the shape of the primordial power spectra could be modified in a much more important way in the Deformed Algebra scheme than in the approach presented in this manuscript. This point definitely requires deeper investigations that are not the purpose of this paper.\\

We have focused on the scalar spectrum which is the more interesting one from the viewpoint of observations (and the more intricate to calculate). The general trends however obviously remain true for the tensor spectrum as the new features are due to the modified background dynamics.

\section{Conclusion}

Quantization ambiguities are unavoidable. The only serious requirement in this framework is to recover general relativity (or the Wheeler-DeWitt equation) in the low energy limit of the theory. This leaves an infinite dimensional set of ambiguities in the choice of the function $f$. Investigating few of its phenomenological properties was the goal of this study.\\

We have shown that a long enough inflationary stage can be generated by an appropriate generalized holonomy correction without the need for matter violating the energy conditions. This however requires a serious amount of fine-tuning in the choice of the parameters entering the definition of the function. Nevertheless this concept is ill-defined when no natural measure is available. More profoundly, it is worthwhile emphasizing that the fine-tuning is a real issue only when it is required to produce a situation which is {\it a priori} singular or specific (in a Bayesian sense). Otherwise, it is just a mere version of the nonproblematic tautology ``if the laws had been different, the World would be different". In addition, we have shown that even though a nearly scale invariant primordial power spectrum can be generated, there are some (often forgotten) associated issues to face, making the spectrum hardly compatible with data.\\

The influence of the shape and amplitude of the holonomy correction on the background dynamics has also been investigated in details. The effects are quite weak and do not change drastically the usual conclusions of LQC as long as the used function remains close enough to the standard case. Interestingly some new effects tend to decrease the number of e-folds. This is a good news for phenomenology. With 140 inflationary e-folds, all the subtle quantum gravitational effects are deeply super-Hubble today. Smaller values bring back the hope that some footprints of the bounce might be seen in the CMB, especially when taking into account nonlinear effects where different modes fail to fully decouple one from the other.

\section{Acknowledgements}

We would like to thank Alejandro Perez who was at the origin of this work and who provided many important inputs and explanations. We also thank Salvatore Ribisi and Lautaro Amedei for valuable discussions.

\bibliography{refs.bib}

 \end{document}